\documentstyle[preprint,aps,floats,epsf]{revtex}

\begin{document}
\tighten

\preprint{\hbox to \hsize{\hfill
\vbox{
\hbox{\bf MADPH-00-1177}
\hbox{\bf UH-511-964-00}
\hbox{\bf VAND-TH-00-5}
\hbox{\bf AMES-HET 00-04}
\hbox{May 2000}}}}

\title{\vspace{.5in}
$CPT$-odd Resonances in Neutrino Oscillations}

\author{V. Barger$^1$, S. Pakvasa$^2$, T.J. Weiler$^3$, and K. Whisnant$^4$}

\address{\vspace{.2in}
$^1$Department of Physics, University of Wisconsin, Madison, WI 53706, USA\\
$^2$Department of Physics and Astronomy, University of Hawaii, Honolulu, HI
96822, USA\\
$^3$Department of Physics and Astronomy, Vanderbilt University, Nashville, TN
37235, USA\\
$^4$Department of Physics and Astronomy, Iowa State University, Ames, IA 50011,
USA}

\maketitle

\begin{abstract}

We consider the consequences for future neutrino factory experiments of
small $CPT$-odd interactions in neutrino oscillations. The
$\nu_\mu\to\nu_\mu$ and $\bar\nu_\mu\to\bar\nu_\mu$ survival
probabilities at a baseline $L=732$~km can test for $CPT$-odd
contributions at orders of magnitude better sensitivity than present
limits. Interference between the $CPT$-violating interaction and
$CPT$-even mass terms in the Lagrangian can lead to a resonant
enhancement of the oscillation amplitude. For oscillations in matter, a
simultaneous enhancement of both neutrino and antineutrino oscillation
amplitudes is possible.

\end{abstract}

\thispagestyle{empty}
\newpage

\section{Introduction}

The discrete symmetries $C$, $P$, and $T$ have fundamental importance in
elementary particle theory. Violations of $C$, $P$, $CP$ and $T$ by the
weak interactions have all been observed\cite{I}. $CPT$ invariance is a
basic property of local quantum field theory\cite{H} and no evidence of
deviations from $CPT$ invariance have been found so far. The most
stringent limit on $CPT$ violation is obtained from the difference of
the $K^0$ and $\bar K^0$ masses\cite{E},
\begin{equation}
m_K - m_{\bar K} < 0.44 \times 10^{-18}\rm\ GeV \,.
\end{equation}
In string theory, the $CPT$ invariance may not be manifest due to the
extended nature of strings\cite{C,D}. Mechanisms by which string
theories could spontaneously break $CPT$ have been
formulated\cite{C,D}. The search for $CPT$ violation is thus of
considerable theoretical interest as a means of searching for purely
string effects. Neutrino oscillations have been considered as phenomena
that could probe $CPT$ non-conservation\cite{A}. With growing interest
in the construction of neutrino factories to make high-precision
measurements of neutrino mass-squared differences and of the $CP$
violating phase in the neutrino sector\cite{J}, it is appropriate to
undertake a more extensive study of the ability to measure
$CPT$-violating effects in neutrino oscillations. We find that a
comparison of $\nu_\mu \to \nu_\mu$ and $\bar\nu_\mu \to
\bar\nu_\mu$ oscillation probabilities at neutrino factories would
give precision tests of $CPT$.
A significant result of our study, as reported below, is that
$CPT$ violating resonance effects can occur that can magnify small $CPT$
violation into a measureable oscillation amplitude.

\section{Basic formalism}

Consequences of $CP$, $T$ and $CPT$ violation for neutrino oscillations
have been written down before\cite{K}. We summarize them briefly for the
$\nu_\alpha\to\nu_\beta$ flavor oscillation probabilities
$P_{\alpha\beta}$ at a distance $L$ from the source. If
\begin{equation}
P_{\alpha\beta}(L) \neq P_{\bar\alpha\bar\beta}(L) \,,
\qquad \beta \ne \alpha \,,
\end{equation}
then $CP$ is not conserved.
If
\begin{equation}
P_{\alpha\beta}(L) \neq P_{\beta\alpha}(L) \,,
\qquad \beta \ne \alpha \,,
\end{equation}
then $T$-invariance is violated.
If 
\begin{eqnarray}
P_{\alpha\beta}(L) &\neq& P_{\bar\beta\bar\alpha}(L)\,,
\qquad \beta \ne \alpha \,,
\\
\noalign{\hbox{or}}
P_{\alpha\alpha}(L) &\neq& P_{\bar\alpha\bar\alpha}(L) \,,
\end{eqnarray}
then $CPT$ is violated.
When neutrinos propagate in matter, matter effects give rise to apparent  
$CP$ {\it and} $CPT$ violation even if the mass matrix is $CP$ conserving.

The $CPT$ violating terms can be Lorentz-invariance violating (LV) or
Lorentz invariant. The Lorentz-invariance violating,
$CPT$ violating case has been discussed by Coleman and Glashow\cite{A}
and by Colladay and Kostelecky\cite{B}. We will consider this first.

The effective LV $CPT$ violating interaction for neutrinos is of the form
\begin{equation}
\bar\nu_L^\alpha b_\mu^{\alpha\beta} \gamma_\mu \nu_L^\beta \,,
\label{eq:LV}
\end{equation}
where $\alpha$ and $\beta$ are flavor indices. We assume rotational
invariance in the ``preferred'' frame, in which the cosmic microwave
background radiation is isotropic (following Coleman and
Glashow~\cite{A}).~\footnote{
An experimental limit on $b^{ee}_3$ of $10^{-29}$~GeV has been
obtained~\cite{M} in studies of torques on a spin polarized torsion
pendulum. This translates into a bound on the $ee$ element of the matrix
$b_0$ of $5\times10^{-25}$~GeV; if $SU(2)_L$ symmetry holds, a similar
bound is implied on the $\nu_e\nu_e$ element of $b_0$, but there are no similar bounds on other $\nu\nu$ elements of $b_0$.}
The energies of ultra-relativistic neutrinos with definite momentum $p$
are eigenvalues of the matrix
\begin{equation}
m^2/2p + b_0 \,,
\end{equation}
where $b_0$ is a hermitian matrix, hereafter labeled $b$.

In the two-flavor case the neutrino phases may be chosen such that
$b$ is real, in which case the interaction in Eq.~(\ref{eq:LV}) is
$CPT$-odd. The survival probabilities for flavors $\alpha$ and
$\bar\alpha$ produced at $t=0$ are given by\cite{A}
\begin{eqnarray}
P_{\alpha\alpha}(L) &=&
1 - \sin^22\Theta \sin^2(\Delta L/4)\,, \label{eq:foo}\\
\noalign{\hbox{and}}
P_{\bar\alpha\bar\alpha}(L) &=&
1 - \sin^2 2\bar\Theta \sin^2(\bar\Delta L/4) \,,\\ 
\noalign{\hbox{where}}
\Delta\sin2\Theta &=&
\left| (\delta m^2/E) \sin2\theta_m
+ 2\delta b e^{i\eta} \sin2\theta_b \right| \,,
\label{eq:delsin}\\
\Delta\cos2\Theta &=&
(\delta m^2/E) \cos2\theta_m + 2\delta b \cos2\theta_b \,.
\label{eq:delcos}
\end{eqnarray}
$\bar\Delta$ and $\bar\Theta$ are defined by similar equations with
$\delta b\to -\delta b$.  Here $\theta_m$ and $\theta_b$ define the
rotation angles that diagonalize $m^2$ and $b$, respectively, $\delta
m^2 = m_2^2 - m_1^2$ and $\delta b = b_2 - b_1$, where $m_i^2$ and $b_i$
are the respective eigenvalues. We use the convention that
$\cos2\theta_m$ and $\cos2\theta_b$ are positive and that $\delta m^2$
and $\delta b$ can have either sign.  The phase $\eta$ in
Eq.~(\ref{eq:delsin}) is the difference of the phases in the unitary
matrices that diagonalize $\delta m^2$ and $\delta b$; only one of these
two phases can be absorbed by a redefinition of the neutrino states.

Observable $CPT$-violation in the two-flavor case is a consequence of
the interference of the $\delta m^2$ terms (which are $CPT$-even) and
the LV terms in Eq.~(\ref{eq:LV}) (which are $CPT$-odd); if $\delta m^2
= 0$ or $\delta b = 0$, then there is no observable $CPT$-violating
effect in neutrino oscillations.
%~\footnote{In the two-flavor case, the matrices that diagonalize the
%mass and $b$ terms are real, and only the relative phase survives. As a
%result, there is no observable $T$ violation for two
%flavors~\cite{L}. For three or more flavors, there are non-trivial
%phases from the diagonalizations, which give rise to observable
%$T$-violating effects in addition to $CPT$ violation.}.
If $\delta m^2/E \gg 2\delta b$ then
$\Theta \simeq \theta_m$ and $\Delta \simeq \delta m^2/E$, whereas if
$\delta m^2/E \ll 2\delta b$ then $\Theta \simeq \theta_b$ and $\Delta
\simeq 2\delta b$. Hence the effective mixing angle and oscillation
wavelength can vary dramatically with $E$ for appropriate values of
$\delta b$.

We note that a $CPT$-odd resonance for neutrinos ($\sin^22\Theta = 1$)
occurs whenever $\cos2\Theta = 0$ or
\begin{equation}
(\delta m^2/E) \cos2\theta_m + 2\delta b \cos2\theta_b = 0\,;
\end{equation}
similar to the resonance due to matter effects~\cite{F,G}. The condition
for antineutrinos is the same except $\delta b$ is replaced by $-\delta
b$. The resonance occurs for neutrinos if $\delta m^2$ and $\delta b$
have the opposite sign, and for antineutrinos if they have the same
sign. A resonance can occur even when $\theta_m$ and $\theta_b$ are both
small, and for all values of $\eta$; if $\theta_m = \theta_b$, a
resonance can occur only if $\eta \ne 0$.

If one of $\nu_\alpha$ or $\nu_\beta$ is $\nu_e$, then the neutrino
propagation is modified in the presence of matter. Then
Eq.~(\ref{eq:delcos}) becomes
\begin{equation}
\Delta\cos2\Theta = (\delta m^2/E) \cos2\theta_m
+ 2\delta b \cos2\theta_b - 2 \sqrt2 G_F N_e
\,,
\label{eq:delcos2}
\end{equation}
for neutrinos, where $N_e$ is the number density of electrons in matter.
For antineutrinos, $\delta b \to -\delta b$ and $N_e \to -N_e$ in
Eq.~(\ref{eq:delcos2}). 

\section{Examples of $CPT$-violation and $CPT$-odd resonances}

Hereafter, for simplicity, we assume that $m^2$ and $b$ are diagonalized
by the same angle $\theta$, i.e., $\theta_m = \theta_b \equiv \theta$.

\subsection{$\eta = 0$}

For $\eta=0$ we have
\begin{eqnarray}
\Theta &=& \theta \,,
\label{eq:tan}\\
\Delta &=& (\delta m^2/E) + 2\delta b \,.
\label{eq:delta}
\end{eqnarray}
In this $\theta_m = \theta_b$, $\eta=0$ case a resonance is not
possible. The oscillation probabilities become
\begin{eqnarray}
P_{\alpha\alpha}(L) &=& 1 - \sin^2 2\theta \sin^2 \left\{ \left( {\delta m^2  
\over 4E} + {\delta b\over 2} \right) L \right\} \,,
\label{eq:P}\\
P_{\bar\alpha\bar\alpha}(L) &=& 1 - \sin^2 2\theta \sin^2 \left\{ \left(  
{\delta m^2 \over 4E} - {\delta b\over 2} \right) L \right\} \,.
\label{eq:Pbar}
\end{eqnarray}
For fixed $E$, the $\delta b$ terms act as a phase shift in the
oscillation argument; for fixed $L$, the $\delta b$ terms act as a
modification of the oscillation wavelength.
%In either case, the effect
%of $\delta b$ on $P_{\bar\alpha\bar\alpha}$ is the opposite of that on
%$P_{\alpha\alpha}$.

An approximate direct limit on $\delta b$ when $\alpha = \mu$ can be
obtained by noting that in atmospheric neutrino data the flux of
downward going $\nu_\mu$ is not  depleted whereas that of upward going
$\nu_\mu$~is\cite{atmos}. Hence, the oscillation arguments in
Eqs.~(\ref{eq:P}) and (\ref{eq:Pbar}) cannot have fully developed for
downward neutrinos. Taking $|\delta b L/2| < \pi/2$ with $L\sim20$~km
for downward events leads to the upper bound $|\delta b| <
3\times10^{-20}$~GeV; upward going events could in principle test
$|\delta b|$ as low as $5\times10^{-23}$~GeV.  Since the $CPT$-odd
oscillation argument depends on $L$ and the ordinary oscillation
argument on $L/E$, improved direct limits could be obtained by a
dedicated study of the energy and zenith angle dependence of the
atmospheric neutrino data.

\begin{figure}
\centering\leavevmode
\epsfxsize=5in\epsffile{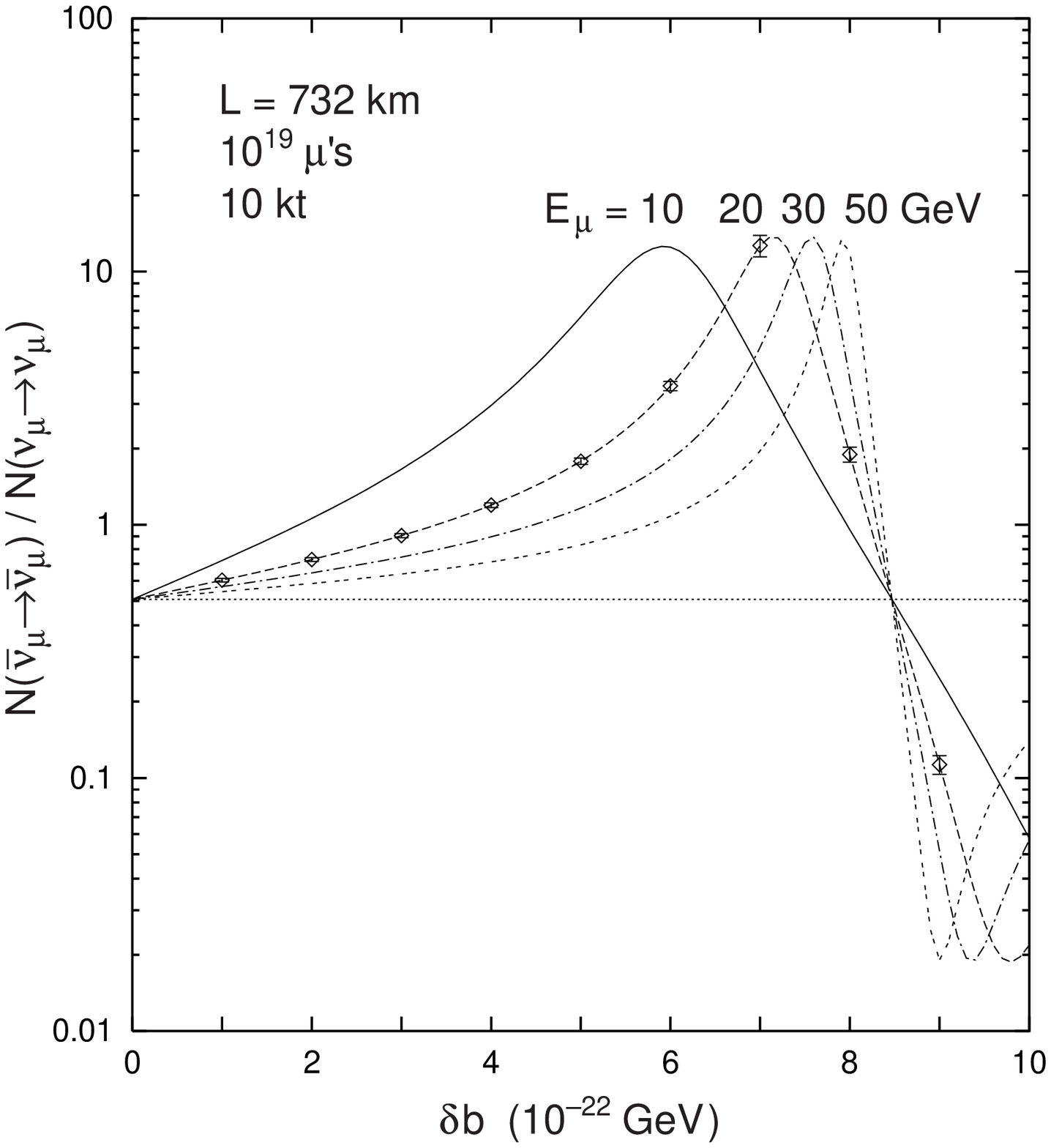}

\bigskip
\caption[]{The ratio of $\bar\nu_\mu\to\bar\nu_\mu$ to
$\nu_\mu\to\nu_\mu$ event rates in a 10~kt detector for a neutrino
factory with $10^{19}$ stored muon with energies $E_\mu = 10$, 20, 30,
50~GeV for baseline $L=732$~km versus the $CPT$-odd parameter $\delta b$
with $\theta_m = \theta_b \equiv \theta$ and phase $\eta=0$. The
neutrino mass and mixing parameters are $\delta m^2 =
3.5\times10^{-3}$~eV$^2$ and $\sin^22\theta = 1.0$. The dotted line
indicates the result for $\delta b = 0$, which is given by the ratio of
the $\bar\nu$ and $\nu$ charge-current cross sections. The error bars
are representative statistical uncertainties.}
\label{fig:ratio}
\end{figure}

The difference between $P_{\alpha\alpha}$ and $P_{\bar\alpha\bar\alpha}$
\begin{equation}
P_{\alpha\alpha}(L) - P_{\bar\alpha\bar\alpha}(L) =
- 2 \sin^22\theta \sin\left({\delta m^2 L\over2E}\right) \sin(\delta b L) 
\,, \label{eq:deltaP}
\end{equation}
can be used to test for $CPT$-violation. In a neutrino factory, the
ratio of $\bar\nu_\mu \to \bar\nu_\mu$ to $\nu_\mu \to \nu_\mu$ events
will differ from the standard model (or any local quantum field theory
model) value if $CPT$ is violated. Figure~\ref{fig:ratio} shows the
event ratios $N(\bar\nu_\mu \to \bar\nu_\mu)/N(\nu_\mu \to \nu_\mu)$
versus $\delta b$ for a neutrino factory with 10$^{19}$ stored muons and
a 10~kt detector at several values of stored muon energy, assuming
$\delta m^2 = 3.5\times10^{-3}$~eV$^2$ and $\sin^22\theta = 1.0$, as
indicated by the atmospheric neutrino data~\cite{atmos}. The error bars
in Fig.~\ref{fig:ratio} are representative statistical
uncertainties. The node near $\delta b = 8\times10^{-22}$~GeV is a
consequence of the fact that $P_{\alpha\alpha} =
P_{\bar\alpha\bar\alpha}$, independent of $E$, whenever $\delta b L = n
\pi$, where $n$ is any integer; the node in Fig.~\ref{fig:ratio} is for
$n=1$. A $3\sigma$ $CPT$ violation effect is possible in such an
experiment for $\delta b$ as low as $3\times10^{-23}$~GeV for stored
muon energies of 20~GeV. Although matter effects also induce an apparent
$CPT$-violating effect, the dominant oscillation here is $\nu_\mu \to
\nu_\tau$, which has no matter corrections in the two-neutrino
limit; in any event, the matter effect is in general small for distances
much shorter than the Earth's radius.

We have also checked the observability of $CPT$ violation at other
distances, assuming the same neutrino factory parameters used above.
For $L=250$~km, the $\delta b L$ oscillation argument in
Eq.~(\ref{eq:deltaP}) has not fully developed and the ratio of $\bar\nu$
to $\nu$ events is still relatively close to the standard model value.
For $L=2900$~km, a $\delta b$ as low as $10^{-23}$~GeV may be observable
at the $3\sigma$ level. However, longer distances may
also have matter effects that simulate $CPT$ violation.

\subsection{$\eta = \pi/2$}

\begin{figure}
\centering\leavevmode
\epsfxsize=5in\epsffile{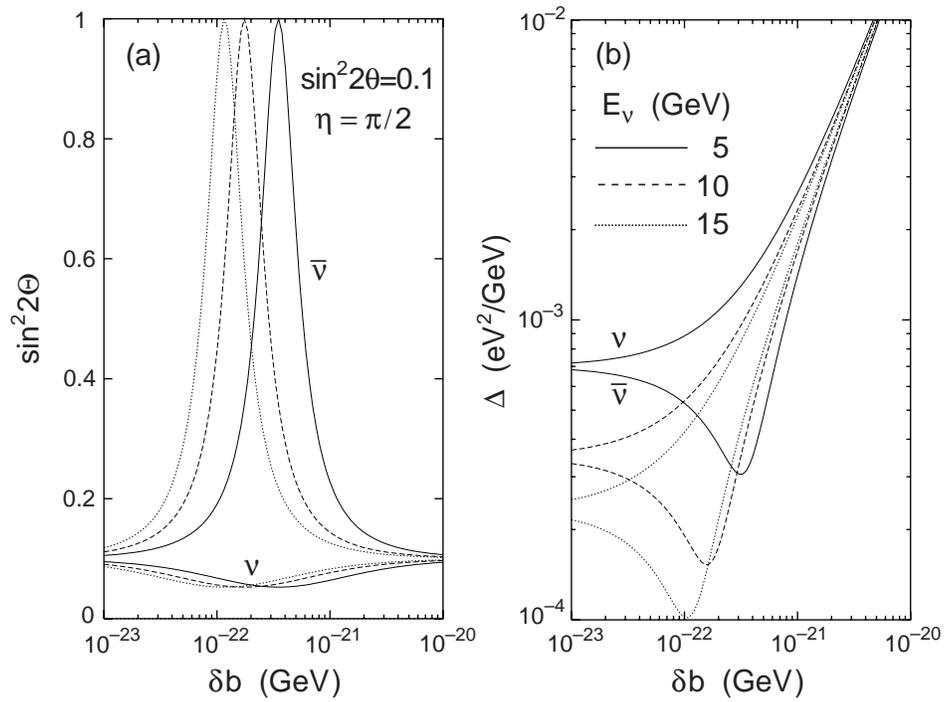}

\bigskip
\caption[]{Resonance effects in $\nu\to\nu$ and $\bar\nu\to\bar\nu$  
oscillations shown versus $CPT$-odd parameter $\delta b$ for various
values of neutrino energy $E$ with $\delta m^2 = 3.5\times
10^{-3}\rm\,eV^2$, $\sin^22\theta = 0.1$ and phase $\eta = \pi/2$:
(a)~oscillation amplitude $\sin^22\Theta$ in Eq.~(\ref{eq:tan2}) and
(b)~oscillation argument $\Delta$ in Eq.~(\ref{eq:delta2}).}
\label{fig:Todd}
\end{figure}

For $\eta = \pi/2$ we have
\begin{eqnarray}
P_{\alpha\alpha} &=&
1 - \sin^2 2\Theta \sin^2 \left\{ \Delta L/4 \right\}\,,
\label{eq:P2}\\
P_{\bar\alpha\bar\alpha} &=& 1 - \sin^2 2\bar\Theta \sin^2 \left\{ \bar\Delta  
L/ 4 \right\} \,,
\label{eq:Pbar2}
\end{eqnarray}
where
\begin{eqnarray}
\tan 2\Theta &=& { \sqrt{ (\delta m^2/E)^2 + (2\delta b)^2 } \over
(\delta m^2/ E) + 2\delta b } \tan2\theta \,,
\label{eq:tan2}\\
\Delta^2 &=& \left[ (\delta m^2/E) + 2\delta b \right]^2
- 4(\delta m^2/E) \delta b \sin^2 2\theta \,,
\label{eq:delta2}
\end{eqnarray}
and $\bar\Theta$ and $\bar\Delta$ are defined similarly with $\delta b
\to -\delta b$. In this case the resonance condition for neutrinos is
$\delta m^2/E + 2\delta b = 0$. Figure~\ref{fig:Todd} shows the effective
oscillation amplitude $\sin^22\Theta$ and oscillation argument $\Delta$
versus $\delta b$ with $\delta m^2 = 3.5\times10^{-3}$~eV$^2$ and
$\sin^22\theta = 0.1$ (which may be appropriate for $\nu_e \to \nu_e$
oscillations) for several values of neutrino energy. Although
the above example assumed $\eta = \pi/2$, such a resonance can occur in this $\theta_b = \theta_m$ example for
any value of $\eta$ in the open interval $(0,2\pi)$.

\subsection{$CPT$-odd term with matter}

In the presence of matter, the effective $\nu_e$ oscillation amplitude and
argument are defined by Eqs.~(\ref{eq:delsin}) and (\ref{eq:delcos2}).
Again assuming $\theta_b = \theta_m \equiv \theta$ and $\eta = 0$, we have
\begin{eqnarray}
\tan 2\Theta &=&
{ \left[ (\delta m^2/E) + 2\delta b \right] \sin 2\theta  \over
\left[ (\delta m^2/E) + 2\delta b \right] \cos 2\theta - 2\sqrt2 G_F
N_e} \,, 
\label{eq:tan3}\\
\Delta^2 &=&
\left\{ \left[ (\delta m^2/E) + 2\delta b \right] \cos2\theta
- 2\sqrt2 G_F N_e \right\}^2 + \left[ (\delta m^2/E) + 2 \delta b
\right]^2 \sin^22\theta \,,
\label{eq:delta3}
\end{eqnarray}
for neutrinos, with $\delta b \to -\delta b$ and $N_e \to - N_e$ for
antineutrinos. Thus a resonance ($\sin^22\Theta = 1$) occurs for
neutrinos when $[(\delta m^2/E) + 2\delta b]\cos2\theta = 2\sqrt2 G_F
N_e$, and for antineutrinos when $[(\delta m^2/E) - 2\delta
b]\cos2\theta = - 2\sqrt2 G_F N_e$. A resonance can occur simultaneously
for neutrinos {\it and} antineutrinos only in the limit when $\delta
m^2/E \ll 2\delta b$ and the $CPT$-odd effects dominate. However, it is
possible to have an effective oscillation amplitude that is
significantly enhanced for both neutrinos and antineutrinos even when
$\delta m^2/E$ is not small compared to $2\delta b$. For $N_e =
1.67 N_A$/cm$^3$ (the electron density appropriate for the upper mantle
of the Earth) and vacuum amplitude $\sin^22\theta = 0.1$, the effective
oscillation amplitudes $\sin^22\Theta$ for $\nu_e \to \nu_e$ and
$\sin^22\bar\Theta$ for $\bar\nu_e \to \bar\nu_e$ can both be greater
than 0.5 when $\delta b$ and $\delta m^2$ satisfy both
.0002~eV$^2$/GeV$ < 2 \delta b + (\delta
m^2/E) < $.0004~eV$^2$/GeV and .0002~eV$^2$/GeV$ < 2 \delta b -
(\delta m^2/E) < $.0004~eV$^2$/GeV. These conditions are satisfied when
$\delta b \simeq 1$--$2\times10^{-22}$~GeV and with $|\delta m^2/E|$ as
large as $10^{-4}$~eV$^2$/GeV. Assuming $\delta m^2 \simeq
3.5\times10^{-3}$~eV$^2$, such enhancements in $\nu_e \to \nu_e$ and
$\bar\nu_e \to \bar\nu_e$ are possible for $E > 35$~GeV, provided that
$\delta b > 0$. Although here we have considered the case $\eta=0$,
similar enhancements are possible for any value of $\eta$ since they
rely on the denominator of Eq.~(\ref{eq:tan3}) being small, which is
independent of $\eta$.

\section{Lorentz-invariant case}

We can simulate a possible Lorentz-invariant $CPT$-odd
effective interaction by allowing the mass matrix for $\bar\nu$'s to be
different from the one for $\nu$'s.
%In this case the survival
%probability for $\nu_\alpha$ and $\bar\nu_\alpha$ will be given by
%
%\begin{eqnarray}
%P_{\alpha\alpha} &=& 1 - \sin^2 2\theta \sin^2
%[\delta m^2 L/(4E)] \,,\\
%P_{\bar\alpha\bar\alpha} &=& 1 - \sin^2 2\bar\theta \sin^2
%[\delta \bar m^2 L/(4E)] \,.
%\end{eqnarray}
%
If we assume, for simplicity, that the $CPT$-violating effects are more  
important in $\delta m^2$ than in mixing, then there is only one $CPT$-odd  
parameter, namely $\delta m^2 - \delta \bar m^2 = \epsilon$, and the
oscillation probabilities are
\begin{eqnarray}
P_{\alpha\alpha} &=& 1 - \sin^2 2\theta \sin^2 [\delta m^2 L/(4E)] \,,
\\
P_{\bar\alpha\bar\alpha} &=& 1 - \sin^2 2\theta \sin^2
[(\delta m^2-\epsilon)L/(4E)] \,.
\end{eqnarray}
{}From the lack of large disappearance of downward going atmospheric
muons, we obtain an approximate upper bound of $|\epsilon|< 0.1$~eV$^2$
when $\alpha = \mu$. A fit to the total number of muon and antimuon
events in the SuperK atmospheric neutrino data sample would greatly
improve this bound.

\section{Summary}

We have shown that small $CPT$-odd interactions of neutrinos can have
measureable consequences in neutrino oscillations. Resonant enhancements
of the oscillation amplitude for either neutrinos or antineutrinos (but
not both) are possible if the unitary matrices which diagonalize the
neutrino mass term and the $CPT$-odd term are not the same. A resonance
can occur for any relative phase between the $CPT$-even mass term and
the $CPT$-odd interaction, but if the rotation angles in the two sectors
are the same, a resonance is possible only if the relative phase is not
zero. In matter, significant enhancements are possible for both
neutrinos and antineutrinos.  Measurement of $\nu_\mu \to \nu_\mu$ and
$\bar\nu_\mu \to \bar\nu_\mu$ oscillation probabilities in neutrino
factories can place stringent limits on the $CPT$-odd interaction.

\section*{Acknowledgments}

The authors would like to thank J.~Learned, J.~Lykken and
X.~Tata for useful discussions. This work was supported in part by the
U.S. Department of Energy, Division of High Energy Physics, under Grants
No.~DE-FG02-94ER40817, No.~DE-FG05-85ER40226, and No.~DE-FG02-95ER40896, and in part by the University of Wisconsin Research Committee with funds granted by the Wisconsin Alumni Research Foundation.

\newpage

\end{document}